
\magnification=\magstep1
\baselineskip=20 pt
\overfullrule=0pt
\def\i{\item}

\def\L{\cal L}
\def\slp{\raise.15ex\hbox{$/$}\kern-.57em\hbox{$\partial$}}
\def\sla{\raise.15ex\hbox{$/$}\kern-.57em\hbox{$a$}}
\def\slA{\raise.15ex\hbox{$/$}\kern-.57em\hbox{$A$}}
\def\slB{\raise.15ex\hbox{$/$}\kern-.57em\hbox{$B$}}
\def\slb{\raise.15ex\hbox{$/$}\kern-.57em\hbox{$b$}}
\def\slf{\raise.15ex\hbox{$/$}\kern-.57em\hbox{$f$}}
\def\slh{\raise.15ex\hbox{$/$}\kern-.57em\hbox{$h$}}
\def\slW{\raise.15ex\hbox{$/$}\kern-.57em\hbox{$W$}}

\def\dda{\sqcup\!\!\!\!\sqcap}
\def\Buildrel#1\over#2{\mathrel{\mathop{\kern0pt #1}\limits_{#2}}}
\quad
\rightline{HD-THEP-95-17}
\vskip1.5cm
\centerline{\bf Bosonisation in Three-Dimensional Quantum Field Theory}
\vskip1cm
\centerline{R. Banerjee\footnote*{On leave of absence
  from S.N.Bose Natl. Ctr. for Basic Sciences, DB-17, Sec. 1, Salt Lake,
Calcutta 700064, India}}
\medskip
\centerline{Institut f\"ur Theoretische Physik}
\centerline{Universit\"at Heidelberg}
\centerline{Philosophenweg 16, D-69120 Heidelberg}
\centerline{Germany}
\vskip2cm
\centerline{\bf Abstract}
\bigskip
We show in three dimensions, using functional integral
techniques, the equivalence between the partition functions
of the massive Thirring model and a gauge theory with two gauge
fields, to \underbar{all} orders in the inverse fermion
mass. Detailed bosonisation identities, also valid to
\underbar{all} orders in the inverse mass, are derived. Specialisation to the
lowest (and next to lowest) orders reveals that the gauge
theory simplifies to the Maxwell-Chern-Simons theory. Some interesting
consequences of the mapping are discussed in this case.
\vfill\eject
The mapping of bosonic theories into fermionic ones and vice
versa, commonly known as bosonisation (fermionisation),
provide a powerful approach to study (in 1+1 dimensions)the non-perturbative
behaviour of either quantum field theories [1] or condensed
matter systems [2]. Although bosonisation is reasonably
well understood in 1+1 dimensional theories [3], the
situation is much less clear in higher dimensions. This
is because a Mandelstam [1]-like operator construction is
non-trivial in higher dimensions. Similarly, Schwinger terms
in the current algebra in these cases is also rather involved
so that bosonisation identities relating the currents in
the different (bosonic or fermionic) theories cannot be so easily
deduced as happens in 1+1 dimensions. Very recently,
however,
Fradkin and Schaposnik [4] have shown the equivalence (in 2+1
dimensions) of the massive Thirring model, to the leading order in
the inverse mass, with the Maxwell-Chern-Simons gauge theory.
This work may be viewed in the perspective of ideas initiated
by Polyakov [5] and elaborated by Deser and Redlich [6], who
had revealed a connection between the 2+1 dimensional $CP^1$ model
with a Chern-Simons term and of a charged massive fermion, to
lowest order in the inverse fermion mass. The extension of
these findings to higher orders is
problematic and even leads to ambiguities [6].

In the present paper we discuss bosonisation in
2+1 dimensions, similar in spirit to [4],
within a path-integral framework. The partition function of
the massive Thirring model (MTM) is shown to be equivalent
to \underbar{all} orders in the inverse fermion mass, with
the partition function of an interacting gauge theory involving
two gauge fields. Detailed bosonisation identities, once again
valid to \underbar{all} orders
in the inverse fermion mass, mapping the Thirring current
and its dual with corresponding (dual) field strengths in the
gauge theory, are provided. A similar mapping for the free
term of the MTM is also given. The results of [4], which are
valid only in the leading order, are reproduced when the MTM
is identified with the Maxwell-Chern-Simons (MCS) theory.
Furthermore, in the next-to-leading order, the MTM is still
shown to be equivalent to the MCS theory but the mass of
the gauge boson is renormalised. Application of the bosonisation identities
in this case show that for the MTM; (i) the current satisfies a
self- duality relation, (ii) there is a Schwinger term in the current algebra
 and (iii) the energy-momentum tensor can be expressed in a Sugawara [7] form.

Consider the following (2+1 dimensional) master Lagrangian,
$${\L}=\bar\psi^i(i\slp-m-{\lambda\over
{\sqrt N}}\slf)\psi^i-{1\over4}
F^2_{\mu\nu}+\epsilon_{\mu\nu\rho}f^\mu\partial^\nu
A^\rho\eqno(1)$$
where $F_{\mu\nu}=\partial_\mu A_\nu-\partial_\nu A_\mu$ and $\psi^i$
are $N$ two-component Dirac spinors. Master Lagrangians [8-10] have
proved useful in exhibiting equivalences between different models
in 2+1 dimensions. In particular, the master Lagrangian suggested
by Deser and Jackiw [8] to reveal the correspondence of a
self-dual model [11] with the MCS theory involved (contrary to
(1))\underbar{only}
Bose fields and was used in [4] to discuss bosonisation. The Lagrangian
(1) is invariant under the independent $U(1)$ gauge transformations,
$$\psi^i\to e^{i\alpha(x)}\psi^i,\ f_\mu\to f_\mu-{\sqrt N\over\lambda}
\partial_\mu\alpha,\ A_\mu\to A_\mu-\partial_\mu\beta\eqno(2)$$
while the equations of motion obtained by varying the different
fields are
$$(i\slp -m-{\lambda\over\sqrt N}\slf)\psi^i=0\eqno(3a)$$
$$j_\mu={\sqrt N\over\lambda}\epsilon_{\mu\nu\lambda}
\partial^\nu A^\lambda={\sqrt N\over\lambda}F_\mu\eqno(3b)$$
$$\epsilon_{\mu\nu\lambda}\partial^\nu f^\lambda=\partial^\nu
F_{\mu\nu}\eqno(3c)$$
where
$$j_\mu=\bar\psi^i\gamma_\mu\psi^i\eqno(4)$$
is the gauge-invariant (conserved) $U(1)$ current and we have
introduced the dual field strength $F_\mu$. The generating
functional for (1) in the presence of external sources $J_\mu,
\tilde J_\mu$ coupled to gauge-invariant fields is given by
$$Z=\int d[\psi,\bar\psi,f_\mu,
A_\mu]\delta(\partial_\mu f^\mu)\delta(\partial_\mu A^\mu)e^{i
\int d^3x[{\L}+F_\mu J^\mu+\tilde f_\mu\tilde J^\mu]}
\eqno(5)$$
where $\tilde f_\mu=\lambda^2\epsilon_{\mu\nu\lambda}\partial^\nu
f^\lambda$ is the dual to $f_\mu$ normalised by $\lambda^2$
so that the sources have identical dimensions, and a covariant
gauge has been chosen for both $f_\mu$ and $A_\mu$ fields.
The Gaussian integration over $A_\mu$ is easily performed by
implementing the gauge $\partial_\mu A^\mu=0$ using 't Hooft's
prescription to yield
$$Z=\int d[\psi,\bar\psi,f_\mu]\delta(\partial_\mu f^\mu)
e^{i
\int d^3x[\bar\psi^i(i\slp-m-{\lambda\over
\sqrt N}\slf)\psi^i+{1\over2}f^\mu f_\mu +f_\mu J^\mu+{1\over2}
J_\mu(g^{\mu\nu}-{\partial^\mu\partial^\nu\over
{\dda}})J_\nu+\tilde f_\mu\tilde J^\mu]}
\eqno(6)$$
We now express $\delta(\partial_\mu f^\mu)$ as a Fourier
transform with variable $\beta(x)$. Then (6) may be written as
$$\eqalign{
&Z=\int d[\psi,\bar\psi,f_\mu,\beta]
e^{i\int d^3x[\bar\psi^i(i\slp-m-{\lambda\over
\sqrt N}(\slf+\slp\beta))\psi^i+{1\over2}(f^\mu+\partial^\mu\beta)
(f_\mu+\partial_\mu\beta)}\cr
&-{1\over2}
\partial_\mu\beta\partial^\mu\beta+(f_\mu+\partial_\mu\beta)
J^\mu+\beta\partial_\mu
J^\mu+{1\over2}J_\mu(g^{\mu\nu}-{\partial^\mu\partial^\nu\over\dda})
J_\nu+\lambda^2\epsilon_{\mu\nu\lambda}\partial^\nu(f^\lambda+\partial
^\lambda\beta)\tilde J^\mu]\cr}\eqno(7)$$
where conservation of the current (3b,4) has been used. (Recall that there
are no divergence anomalies in arbitrary odd-dimensions [12].)
Introducing the new fields $h_\mu=f_\mu+\partial_\mu\beta$
and performing the $\beta$ integration yields
$$Z=\int d[\psi,\bar\psi,h_\mu]e^{i\int d^3x[\bar\psi^i(i\slp
-m-{\lambda\over\sqrt N}\slh)\psi^i+{1\over2}h^\mu h_\mu+h_\mu(J^\mu
+\lambda^2\epsilon^{\mu\nu\lambda}\partial_\nu\tilde J_\lambda)+
{1\over2}J^\mu J_\mu]}\eqno(8)$$
Finally, integrating over $h_\mu$ leads to
$$Z=\int d[\psi,\bar\psi]e^{i\int d^3x[{\L}_{MTM}+
{\lambda\over\sqrt N}j_\mu J^\mu+{\lambda\over\sqrt N}\tilde j_\mu\tilde
J^\mu-{\lambda^4\over 4}(\partial_\mu\tilde J_\nu-\partial_\nu
\tilde J_\mu)^2-\lambda^2\epsilon_{\mu\nu\rho}J^\mu\partial^\nu
\tilde J^\rho]}\eqno(9a)$$
where
$${\L}_{MTM}=\bar\psi^i(i\slp-m)\psi^i-{\lambda^2\over 2N}j_\mu j^\mu
\eqno(9b)$$
is the Lagrangian for the massive Thirring model (MTM) (with $j_\mu$
given in 4), which is known to be renormalisable in the
$1/N$ expansion [13], while
$$\tilde j_\mu=\lambda^2\epsilon_{\mu\nu\lambda}\partial^\nu j^\lambda
\eqno(10)$$
is the dual Thirring current. In the absence of sources $J_\mu,
\tilde J_\mu$ we see that (9) represents the partition function
for the MTM.

Alternatively, doing the fermionic integration in (5) just amounts
to evaluating the fermion determinant in presence of the external
field $f_\mu$. This is a well-known [6,12,14] gauge-invariant expression
computed in inverse powers of the fermion mass which, when substituted
in (5) yields
$$\eqalign{
Z&=\int d[f_\mu,A_\mu]\delta(\partial_\mu f^\mu)\delta(\partial_\mu
A^\mu)e^{i\int d^3x[-{1\over4}F^2_{\mu\nu}+
\epsilon_{\mu\nu\lambda}f^\mu\partial^\nu A^\lambda-}\cr
&\phantom{x}^{-{\lambda^2\over 8\pi}
\epsilon_{\mu\nu\lambda}f^\mu\partial^\nu
f^\lambda+{\lambda^2\over 24\pi m}(\partial_\mu f_\nu-\partial_\nu
f_\mu)^2+0({1\over m^2}\epsilon_{\mu\nu\lambda}f^\mu\partial^\nu
\dda f^\lambda)+F_\mu J^\mu+\tilde f_\mu \tilde J^\mu]}\cr}\eqno(11)$$
In the absence of sources, (11) represents an interacting gauge
theory involving two fields $f_\mu$ and $A_\mu$. Since (9) and (11)
were derived from a common origin (5), we conclude the equivalence of
the partition functions associated with the MTM and a gauge theory
(GT), i.e.
$$Z_{MTM}=Z_{GT}\eqno(12)$$
where $Z_{GT}$ is the r.h.s. of (11) in the absence of sources.
Note that $Z_{GT}$ in the lowest (up to $m^{-1}$) order is just the master
 expression considered in [10] to discuss the connection [8,15] between
the self-dual model of [11] and the MCS theory.

 It is now possible to deduce the bosonisation identities by comparing
the source terms appearing in (9) and (11). We find, modulo
non-propagating contact terms, the following mappings
$$j_\mu\leftrightarrow {\sqrt N\over\lambda}F_\mu={\sqrt N\over\lambda}
\epsilon_{\mu\nu\rho}\partial^\nu A^\rho\eqno(13a)$$
$$\tilde j_\mu\leftrightarrow {\sqrt N\over\lambda}\tilde f_\mu
=\left({\sqrt N\over\lambda}\right)\lambda^2
\epsilon_{\mu\nu\rho}\partial^\nu f^\rho\eqno(13b)$$
where $j_\mu$ and $\tilde j_\mu$ are defined in (4) and (10),
respectively. Thus gauge invariance (and conservation) of the Thirring current
 and its dual are preserved. The above relations are manifestations of the
equations of
motion (3b,c). This happens because of the
Gaussian nature of the problem. Finally, observe that (13a) leads
to the identification of the current-current interaction term in
(9b) with the conventional Maxwell term in (11). This allows the
further correspondence
$$\bar\psi^i(i\slp-m)\psi^i\leftrightarrow \epsilon_{\mu\nu\lambda}
f^\mu\partial^\nu A^\lambda-{\lambda^2\over 8\pi}\epsilon_{\mu\nu\lambda}
f^\mu\partial^\nu f^\lambda+{\lambda^2\over 24\pi m}(\partial_\mu
f_\nu-\partial_\nu f_\mu)^2+0\left({1\over m^2}\right)\eqno(14)$$
Equations (12-14) comprise some of the crucial results of this
paper. Observe that the identification implied by these equations
holds to \underbar{all} orders in the inverse fermion mass.

It is now simple to reproduce the results in [4] valid to the leading
order (i.e. up to ${1\over m}$) in the inverse fermion mass. Note
that the Gaussian nature of (11) allows the possibility of performing
either the $f_\mu$ or $A_\mu$ integration. Doing the integration
over $A_\mu$ does not yield a physically interesting theory (although
 in the leading order, one can show [10] that it yields the self-dual
 model of [11]). We thus
perform the $f_\mu$ integration. Since the Seeley coefficients
expressing the fermion determinant in derivatives of $f_\mu$
(see (11)) are known only up to $0(m^{-2})$ [6], our subsequent
discussion is valid to this order. Implementing the gauge $\partial_\mu
f^\mu=0$ by 't Hooft's method and doing the $f_\mu$ integration,
one finds up to $0(m^{-2})$
$$\eqalign{
Z=&\int DA_\mu\delta(\partial_\mu A^\mu)
e^{i\int d^3 x[-{1\over 4}F^2_{\mu\nu}+{2\pi\over 3m\lambda^2}
F^2_{\mu\nu}+{2\pi\over\lambda^2}\epsilon_{\mu\alpha\beta}A^\mu
\partial^\alpha A^\beta}\cr
&\phantom{x}^{+F_\mu J^\mu+(4\pi F_\mu-{8\pi\over 3m}\partial^\nu F_{\nu\mu})
\tilde J^\mu]}\cr}\eqno(15)$$
where a non-propagating contact term has been dropped. Scaling
 the $A_\mu$ field
to recast the Maxwell term in its conventional form, we find
$$Z=\int DA_\mu\delta(\partial_\mu A^\mu)e^{i\int d^3 x[{\L}_{MCS}+
\left(1+{4\pi\over 3m\lambda^2}\right)
F_\mu(J^\mu +4\pi \tilde J^\mu)-{8\pi\over 3m}\partial^\nu F_{\nu\mu}\tilde
J^\mu]}\eqno(16a)$$
where
$${\L}_{MCS}=-{1\over4}F^2_{\mu\nu}+{M\over 2}\epsilon_{\mu\nu\lambda}
A^\mu\partial^\nu A^\lambda;\ M={4\pi\over\lambda^2}
\left(1+{8\pi\over3m\lambda^2}\right)\eqno(16b)$$
is the MCS Lagrangian where $M$ is the mass of the vector gauge
boson satisfying the equation of motion [16],
$$F_\mu={1\over M}\epsilon_{\mu\nu\lambda}\partial^\nu F^\lambda\eqno(17)$$
expressed in terms of the dual (3b). Thus, up to $0(m^{-2})$, we conclude
$$Z_{MTM}\approx Z_{MCS}\eqno(18)$$
In the leading (up to $m^{-1}$) order this equivalence was
demonstrated in [4]. The mass of the MCS gauge boson found there
was $4\pi/\lambda^2$ which agrees with the leading order term in
$M$ given in (16b).

Comparing the source terms in (9) and (16) reveals the following
bosonisation relations between the gauge-invariant operators in the MTM
(written on the l.h.s.) and MCS theory (written on the r.h.s.)
$$j_\mu\leftrightarrow {\sqrt N\over\lambda}
\left(1+{4\pi\over 3m\lambda^2}\right)F_\mu\eqno(19a)$$
$$\tilde j_\mu=\lambda^2\epsilon_{\mu\nu\lambda}\partial^\nu j^\lambda
\leftrightarrow {4\pi\sqrt N\over\lambda}
\left(1+{4\pi\over3m\lambda^2}\right)
F_\mu-{8\pi\sqrt N\over 3m\lambda}\partial^\nu F_{\nu\mu}\eqno(19b)$$
Using (19a) to recast the current-current interaction term in (9)
as a Maxwell piece that may be identified with (16b) leads to
the further mapping,
$$\bar\psi^i(i\slp-m)\psi^i\leftrightarrow{2\pi\over\lambda^2}
\left(1+{8\pi\over3m\lambda^2}\right)\epsilon_{\mu\nu\lambda}
A^\mu\partial^\nu A^\lambda+{2\pi\over3m\lambda^2}F^2_{\mu\nu}\eqno(20)$$
Equations (18)-(20) constitute the bosonisation identities valid for
an expansion in the inverse fermi mass up to $0(m^{-2})$. These
are the analogues of (12)-(14) that were obtained for all orders, corresponding
 to the gauge theory (11).

Some interesting consequences follow from (19). Notice that
both terms in the r.h.s. of (19b) can be directly expressed
in terms of $j_\mu$ by using (19a). Thus a self-duality relation
for the Thirring current is found,
$$j_\mu={\lambda^2\over 4\pi}\left(1-{8\pi\over 3m\lambda^2}
\right) \epsilon_{\mu\nu\lambda}\partial^\nu j^\lambda\eqno(21)$$
If we substitute back $F_\mu$ in place of $j_\mu$ using (19a),
the MCS equation of motion (17) is reproduced. This serves as
consistency on the overall bosonisation program developed in
the inverse fermion mass.

The canonical (equal time) current algebra is next derived. Using
(19) we see that this algebra just corresponds to the algebra among
the electric and magnetic fields in the MCS theory. Since the latter
is known [16], one can immediately compute the relevant algebra,
$$\eqalign{
&i[j_0(x),j_0(y)]=0\cr
&i[j_0(x),j_i(y)]=-{N\over\lambda^2}\left(1+{8\pi\over 3m\lambda^2}
\right)\partial_i\delta(x-y)\cr
&i[j_l(x),j_m(y)]={4\pi N\over\lambda^4}\left(1+{16\pi\over 3m\lambda^2}
\right)\epsilon_{lm}\delta(x-y)\cr}\eqno(22)$$
We thereby find the existence of a Schwinger term whose structure
is reminiscent of the 1+1 dimensional model. It is easy
to verify that (22) is compatible with the self-duality relation (21).

 Let us next compute the energy momentum
tensor in the MTM (valid up to $0(m^{-2})$) by recalling that the corresponding
 tensor in MCS theory is given in terms of $F_\mu$ by [8,16]
$$\Theta_{\mu\nu}^{MCS} = F_\mu F_\nu -
{g_{\mu\nu}\over 2}F_\beta F^\beta \eqno(23)$$
Using (19a) we immediately obtain the expression for MTM
$$\Theta_{\mu\nu}^{MTM}={\lambda^2\over
N}\left(1-{{8\pi}\over{3m\lambda^2}}\right)
(j_\mu j_\nu - {g_{\mu\nu}\over 2}j_\beta j^\beta)\eqno(24)$$
which, interestingly, has the Sugawara [7] structure. The ``quantum'' nature
of the above construction is manifested by the normalisation which is just
 the inverse of the normalisation in the Schwinger term (22). Exactly the same
 phenomenon occurs in 1+1 dimensions [3] where the equivalence of the Sugawara
 form with the conventional (Noether) energy momentum tensor has also been
 explicitly shown [17] using operator product expansion techniques.
Furthermore,
using (22), the conservation of the current $j_\mu$, as well as the self-dual
 relation (21) follow from (24), as expected. Our findings show that (19) and
(21) may truly be regarded as
``analogues'' of corresponding mappings $j_\mu\sim\epsilon_{\mu\nu}
\partial^\nu\phi,\ j_{\mu5}\sim\partial_\mu\phi,\
j_\mu\sim\epsilon_{\mu\nu}j^\nu_5$
known to exist in 1+1 dimensions.

To conclude, we have shown (in 2+1 dimensions) by starting from a
novel master Lagrangian containing both Bose and Fermi fields,
the equivalence, on the level of partition
functions, of the massive Thirring model (MTM) and a gauge
theory with \underbar{two} gauge fields,
to \underbar{all} orders in the
inverse fermion mass. Bosonisation identities
(on the level of correlation functions) once again valid to
\underbar{all} orders, mapping the current (and its dual)
in the MTM to dual field strengths (corresponding to the
two gauge fields) in the gauge theory are deduced. A similar
identification is also found for the free part of the MTM.
We next specialise up to $0(m^{-2})$ (i.e. the
order up to which the Seeley coefficients in the expansion
of the fermion determinant are known [6](see 11)) computations
whereby one of the gauge fields can be eliminated. To this order
the MTM gets identified with the Maxwell-Chern-Simons (MCS)
theory whose spectrum has a vector gauge boson with mass
${4\pi\over\lambda^2}\left(1+{8\pi\over 3m\lambda^2}\right)$.
In the leading (i.e. up to $m^{-1}$) order this
reproduces the result derived in [4]. Detailed bosonisation
identities (up to $0(m^{-2})$) mapping different operators in the MTM with
those
in the MCS theory are dervied. A self-duality relation for the
Thirring current is found. The current algebra in MTM, computed
from the bosonisation identities,
reveals a Schwinger term. These identities are next used to
 show that the energy momentum tensor in the MTM has a Sugawara [7] form.
 The various bosonisation relations and their consequences
have a striking resemblance with the 1+1-dimensional
case but, contrary to that example, are valid for length scales
long compared to the Compton wavelength of the fermion. Finally,
note that more detailed computations for higher orders (i. e. up
to $0(m^{-3})$) onwards) can be done, exactly as shown here up
to $0(m^{-2})$, provided the relevant Seeley coefficients in
(11) are explicitly known.
 \bigskip I thank H. J. Rothe for fruitful discussions and the Alexander
von Humboldt Foundation for providing financial support.
\bigskip
\centerline{\bf REFERENCES}
\i{[1]} S.Coleman, Phys. Rev.{\bf D11} (1975) 2088; S.Mandelstam,
ibid 3026
\i{[2]} E.Lieb and D.Mattis, Jour. Math. Phys. {\bf 6} (1965) 304; A.Luther
and I.Peschel, Phys. Rev. {\bf B12} (1975) 3908
\i{[3]} For recent reviews see, E.Abdalla, M.Abdalla and K.Rothe,
``Non-perturbative
Methods in 2-Dimensional Quantum Field Theory'' (World Scientific 1991);
E.Fradkin, ``Field Theories of Condensed Matter Systems'' (Frontiers in Physics
1991)
\i{[4]} E.Fradkin and F.Schaposnik, Phys. Lett.{\bf B338} (1994) 253
\i{[5]} A.Polyakov, Mod. Phys. Lett.{\bf A3} (1988) 325
\i{[6]} S.Deser and A.Redlich, Phys. Rev. Lett. {\bf 61} (1988) 1541
\i{[7]} H.Sugawara, Phys. Rev. {\bf 170} (1968) 1659
\i{[8]} S.Deser and R.Jackiw, Phys. Lett. {\bf B139} (1984) 371
\i{[9]} A.Karlhede, U.Lindstrom, M.Rocek and P.van Nieuwenhuizen, Phys.
Lett.{\bf B186} (1987) 96
\i{[10]} R.Banerjee and H.J.Rothe, Heidelberg University Report No.
HD-THEP-95-16
\i{[11]} P.K.Townsend, K.Pilch and P.van Nieuwenhuizen, Phys. Lett. {\bf B136}
(1984) 38
\i{[12]} L.Alvarez-Gaume, S.Della Pietra and G.Moore, Ann. Phys.{\bf 163}
(1985) 288; R.Banerjee, Mod. Phys. Lett.{\bf A6} (1991) 1915
\i{[13]} D.Gross in: Methods in Field Theory, Eds. R.Balian and J.Zinn-Justin,
(North-Holland, Amsterdam, 1976)
\i{[14]} A.Redlich, Phys. Rev.{\bf D29} (1984) 2366;
R.Gamboa-Saravi, M.Muschietti, F.Schaposnik and J.Solomin, Jour. Math. Phys.
{\bf 26} (1985) 2045; K.Babu, A.Das and P.Panigrahi, Phys. Rev. {\bf D36}
(1987) 3725
\i{[15]} R.Banerjee, H.Rothe and K.Rothe, Heidelberg University Report No.
HD-THEP-95-13
\i{[16]} S.Deser, R.Jackiw and S.Templeton, Phys. Rev. Lett. {\bf 48} (1982)
975; Ann. Phys. {\bf 140} (1982) 372
\i{[17]} S.Coleman, D.Gross and R.Jackiw, Phys. Rev.{\bf 180} (1969) 1359
\end